\documentclass[letterpaper,11pt]{article}
\usepackage{bm,graphicx,float,amssymb,amsmath,xcolor,subfigure}
\usepackage[english]{babel}
\usepackage{booktabs}
\usepackage{subcaption}
\usepackage{float}
\usepackage{tikz}
\usepackage{array}
\usepackage[utf8]{inputenc}
\usepackage[title]{appendix}
\usepackage{algorithm}
\usepackage{enumerate}
\usepackage{tabularx}
\usepackage{siunitx}
\usepackage{threeparttable}
\usepackage[noend]{algpseudocode}
\usepackage[round]{natbib}
\usepackage{pgfplots}
\usepackage{enumitem}
\definecolor{myblue}{RGB}{199,229,236}
\definecolor{myorange}{RGB}{245,180,111}
\pgfplotsset{width=10cm,compat=1.18}
\usepackage[colorlinks=true,citecolor=blue]{hyperref}
\newcommand{\sym}[1]{\ifmmode^{#1}\else\(^{#1}\)\fi}
\begin{document}
\title{How FinTech affects financial sustainability: Evidence from Chinese commercial banks using a three-stage network DEA-Malmquist model}
    
\author{
	\normalsize Yudi Yang~\textsuperscript{a}, Fan Yang~\textsuperscript{a,~$*$} , Xiajie Yi~\textsuperscript{b,~$*$}, Dongwei He~\textsuperscript{a,~}\footnote{Corresponding authors. E-mail address:~fan\_yang@shnu.edu.cn (F. Yang), x.yi@ieseg.fr(X. Yi), \\donwayho@shnu.edu.cn (D.~He).} \\
  \\ \small\em \textsuperscript{a} School of Finance and Business, Shanghai Normal University, Shanghai, China
  \\ \small\em  \textsuperscript{b} I{\'E}SEG School of Management, University Lille, CNRS, UMR 9221 - LEM - Lille  \\ \small\em Economie Management, F-59000 Lille, France
  }
\date{}
\maketitle

	\baselineskip 0.8cm
	\begin{abstract}
		\baselineskip 0.8cm
		\noindent 
        This paper investigates the impact of financial technology (FinTech) on the financial sustainability (FS) of commercial banks. We employ a three-stage network DEA-Malmquist model to evaluate the FS performance of 104 Chinese commercial banks from 2015 to 2023. A two-way fixed effects model is utilized to examine the effects of FinTech on FS, revealing a significant negative relationship. Further mechanistic analysis indicates that FinTech primarily undermines FS by eroding banks' loan efficiency and profitability. Notably, banks with more patents or listed status demonstrate greater resilience to FinTech disruptions. These findings help banks identify external risks stemming from FinTech development, and by elucidating the mechanisms underlying FS, enhance their capacity to monitor and manage FS in the era of rapid FinTech advancement.	\end{abstract}

\newpage
\section{Introduction}
\label{sec:Introduction}
Commercial banks play a central role in a financial system by channeling funds from savers to businesses and supporting liquidity creation. Their stability and sustainability are vital to overall economic development. In 2023, Silicon Valley Bank (SVB) went bankrupt due to bank runs, in which many depositors withdrew their deposits simultaneously out of concern that the bank might become insolvent. The collapse of SVB stemmed from an overreliance on short-term deposits from tech firms and excessive exposure to long-term fixed-income securities. When interest rates rose sharply, the bank faced an acute liquidity crisis~\citep{metrick2024failure}. As the 16th largest bank in the US, the collapse of SVB triggered the failures of Silvergate Bank and Signature Bank, raising concerns about the resilience of the banking system and the potential for contagion across the broader financial sector. \cite{aharon2023too} document abnormal returns on both event and post-event days, indicating a negative response to the collapse of SVB in equity markets.~\cite{akhtaruzzaman2023did} indicate that the collapse of SVB severely shook and threatened the global banking industry, and led to a pronounced elevation of financial contagion confined to the banking sector. Overall, this event reveals weaknesses in the risk management of banks and asset allocation. It also underscores the critical role of maintaining financial resilience and sustainability in the banking sector to prevent global systemic risks amid rising interest rates and increasing uncertainty.

\begin{figure}[h!]
\centering
\begin{tikzpicture}
\begin{axis}[
    width=13cm,
    height=9cm,
    ybar=0pt,
    bar width=14pt,
    enlarge x limits=0.12,
    ylabel={Trillion yuan},
    xlabel={Year},
    symbolic x coords={2015,2016,2017,2018,2019,2020,2021,2022,2023},
    xtick=data,
    ymin=0,
    legend style={
        at={(0.5,-0.18)},
        anchor=north,
        legend columns=-1,
        font=\small
    },
    axis x line*=bottom,
    axis y line*=left,
    nodes near coords,
    nodes near coords align={vertical},
    every node near coord/.append style={
        font=\scriptsize,
        anchor=south, 
        black,
    },
    point meta={y}, 
    every node near coord/.append style={
        /pgf/number format/fixed,
        /pgf/number format/precision=2
    },
]
\addplot+[fill=myblue] coordinates {
    (2015,156) (2016,182) (2017,197) (2018,210)
    (2019,239) (2020,266) (2021,289) (2022,320) (2023,355)
};
\addplot+[fill=myorange] coordinates {
    (2015,144) (2016,169) (2017,182) (2018,193)
    (2019,220) (2020,245) (2021,265) (2022,294) (2023,327)
};
\legend{Total assets, Total liabilities}
\end{axis}
\end{tikzpicture}
\caption{The total assets and liabilities of Chinese commercial banks from 2015 to 2023}
\label{fig:assets_liabilities}
\end{figure}
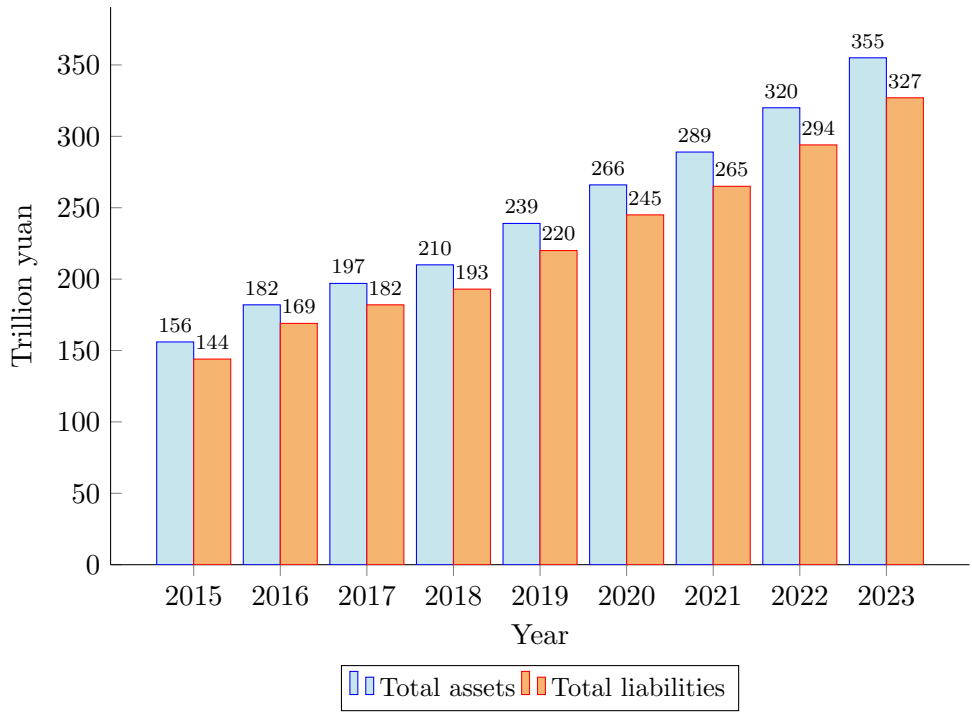

According to the National Administration of Financial Regulation (NAFR)\footnote{The data on total assets and liabilities refer to quarterly figures for legal-person banking institutions. Before 2018, the data were released by the former China Banking Regulatory Commission (CBRC). Since 2023, they have been published by the NAFR, which succeeded the CBRC.}, the total assets of Chinese commercial banks increased from 155.83 trillion yuan in 2015 to 354.85 trillion yuan in 2023, while total liabilities rose from 144.27 trillion yuan to 327.15 trillion yuan during the same period. These numbers represent increases of 127.72\% and 126.76\%, respectively, underscoring the sustained expansion of scale and financial depth of the banking sector, as shown in Figure~\ref{fig:assets_liabilities}. It should be noted that the growth in total assets and liabilities does not necessarily indicate improved financial sustainability (FS). Larger balance sheets, although indicative of business expansion, do not guarantee enhanced profitability or sufficient internal revenue to cover operating costs. At the same time, a rapid increase may entail higher leverage or riskier investments, which can exacerbate costs and financial vulnerability. Here the FS refers to the ability of a firm to cover its operating costs through internal revenue without external support~\citep{zeller2002triangle}. It encompasses various aspects of bank operations, such as financial performance, risk management, and strategic planning. This definition has been widely studied in empirical studies of microfinance institutions (MFIs). One stream of research provides systematic reviews of FS in MFIs~\citep{gupta2023literature, maeenuddin2023developing}. Another stream examines factors affecting FS, including the trade-offs between social objectives and financial performance in MFIs~\citep{wry2018taking}, internal arrangements that influence sustainability~\citep{dabi2023capital}, and analyses of FS in commercial banks, such as its classifications and key determinants~\citep{shi2025bank}. These studies provide a detailed account of the significance of FS in financial institutions, and imply that appropriate indicators must be selected based on its definition.

Measuring FS is an important methodological challenge. Given its multidimensional nature, identifying appropriate measurement methods is essential. Such methods are generally categorized into single-factor and multi-factor ways. The former methods focus on one representative financial ratio to reflect the financial condition of an entity. For example, indicators such as financial self-sufficiency~\citep{kinde2012financial}, return on assets (ROA)~\citep{najam2022towards}, and operational self-sufficiency~\citep{bogan2012capital,  abu2022financial, fonchamnyo2023capital} are commonly used. Single-factor approaches often fail to account explicitly for risk, which may lead to biased assessments of the actual performance of banks~\citep{prior2019profit}. To overcome this limitation, multi-factor measurements often incorporate financial and non-financial variables to capture a more comprehensive picture of FS. Data Envelopment Analysis (DEA)~\citep{charnes1978measuring} is one of the most widely used approaches~\citep{shi2020data}. DEA is a non-parametric method that evaluates organizational performance by integrating multiple input and output indicators into a composite index. The method calculates the efficiency in converting consumed resources into final outputs. Its advantages include not requiring assumptions about the underlying distribution of performance variables and accommodating conflicting performance measures. Consequently, it has been applied in diverse fields such as energy management~\citep{wei2024evaluation}, environmental governance~\citep{zhang2020impact}, public services~\citep{kohl2019use}, the pharmaceutical industry~\citep{qiu2023can}, etc. In the banking industry, scholars have developed multi-stage network DEA models to reflect the structure of the banking sector. For instance, \cite{wang2014efficiency} use an additive two-stage network DEA for Chinese commercial banks to assess overall and stage efficiencies in deposit producing and profit earning stages. Similarly, a two-stage network DEA model is applied by \cite{fukuyama2020nerlovian} to divide the banking process into fund-raising and revenue-generation stages. \cite{shi2025bank} adopt a three-stage network DEA approach for US commercial banks, decomposing the process into deposit, loan, and profitability stages. Overall, the indicators developed in these studies reflect FS across multiple operational stages and show that it is influenced by multiple factors. However, with technological changes, competitive pressures arise not only from internal operations but also from external factors. Among the emerging external factors, the rapid development of financial technology (FinTech) has attracted increasing attention. It is changing the business landscape for banks, expanding their services, and ``interfering'' in the fields traditionally covered by banks~\citep{romanova2016banking}.

Generally, FinTech uses technology to provide new and improved financial services~\citep{thakor2020fintech}. Among the major economies in the world, China offers an example of how FinTech has rapidly taken hold in practice. In 2016, the total financing of Chinese FinTech companies reached 7.7 billion US dollars, surpassing that of the US for the first time, making China the largest market in the world~\citep{gao2022comparison}. By 2022, in terms of transaction volume, Alipay and WeChat Pay accounted for over 90\% of the mobile payment market. The emergence of FinTech exerts various impacts on traditional financial sectors, and its influence on conventional commercial banks remains debated. Existing studies find that FinTech has driven the transformation of the banking sector and challenged traditional banking practices~\citep{romanova2016banking,thakor2020fintech}. On the one hand, FinTech improves efficiency and promotes innovation. \cite{lee2021does} show that FinTech enhances the cost efficiency and technological capabilities of banks by optimizing resource allocation, reducing operational costs, and expanding service boundaries. \cite{wang2021can} also find that commercial banks adopt FinTech to upgrade traditional business models, which helps improve operational efficiency and overall competitiveness. On the other hand, FinTech exerts disruptive effects on banks. Studies show that FinTech diverts business from traditional channels through third-party payments and internet wealth management, thereby reducing interest margins and increasing risks to profitability~\citep{murinde2022impact, lee2023fintech}. \cite{li2023fintech} find that the ``Matthew Effect" in FinTech investment concentrates resources in large banks, placing small and medium-sized banks at a technological and capital disadvantage. Excessive reliance on external technology may also lead to technological dependency and worsen operational risks such as data security and privacy breaches. Furthermore, \cite{wang2021fintech} show that FinTech development increases bank risk-taking, with more potent effects for larger banks with greater involvement in shadow banking. As noted above, the rise of FinTech has introduced both opportunities and threats to banking operations, which affect the FS of banks in turn.

FinTech might promote bank development through technology spillover effects. It may also undermine traditional operations through competition effects. The effect of FinTech on the long-term FS of banks, however, remains unclear. To the best of our knowledge, previous studies do not systematically examine the relationship between FS and FinTech. Therefore, we conduct an empirical analysis to investigate how FinTech influences the FS of Chinese commercial banks. The analysis is based on panel data from 104 commercial banks between 2015 and 2023. Besides, we also perform robustness checks to validate our findings.

The main contributions of this paper are threefold. First, we construct a three-stage network DEA-Malmquist framework to measure the dynamic FS of Chinese commercial banks. This framework also captures the efficiencies of FS in its the deposit, loan, and profitability sub-stages. Second, we use a two-way fixed effects model to examine how FinTech affects FS and the transmission channels, which fills a gap in the literature on FS. Third, we further explore heterogeneity across banks with different levels of innovation and marketization.

The remainder of the paper proceeds as follows. Section~\ref{sec:data} introduces the three-stage network DEA-Malmquist model for the FS and the empirical model. In Section~\ref{sec:analysis}, we empirically evaluate the impact of FinTech on the FS of 104 Chinese commercial banks. Finally, in Section~\ref{sec:Conclusions}, we discuss the findings and future research possibilities.

\section{Methodology}
\label{sec:data}
This section outlines the methodological framework and describes the data used for the regression analysis. To empirically measure FS, we adopt a three-stage network DEA model of \cite{shi2025bank} in combination with the Malmquist productivity index (MI)~\citep{malmquist1953index}. Thereafter, to analyze the relationship between FinTech and FS, we employ a two-way fixed effects model~\citep{wooldridge2010econometric, hsiao2014analysis} and present an overview of the relevant data. The dataset consists of panel data from 104 commercial banks in China from 2015 to 2023. Bank-level data are obtained from the CSMAR\footnote{CSMAR (China Stock Market \& Accounting Research Database), developed by GTA Information Technology Co., Ltd., is widely used for empirical studies on the Chinese capital market.} and Wind\footnote{Wind Database, developed by Wind Information Co., Ltd., provides comprehensive financial and economic data for China.} databases, with missing values supplemented by annual reports\footnote{The reports are derived from the official websites of various banks and the China Foreign Exchange Trade System.} of the commercial banks. Additionally, macroeconomic data are collected from the \textit{China Statistical Yearbook}.

\subsection{The three-stage network DEA-Malmquist model}
\cite{sherman1985bank} first apply DEA to banking studies, after which DEA has been widely used to address banking problems. \cite{staub2010evolution} develop a DEA model to measure cost, technical, and allocative efficiencies of Brazilian banks, finding that inefficiency was mainly technical and state-owned banks were more cost efficient than other types of banks. \cite{avkiran2015illustration} uses a dynamic network DEA model (DN-DEA) to evaluate Chinese commercial banks, showing that DN-DEA captures dynamic performance and highlights sub-unit inefficiencies. Recently, \cite{shi2025bank} construct the FS of commercial banks in the US using a three-stage network DEA model, conceptualizing FS as a multi-stage, multi-factor structure. They also develop a random forest model with SHapley Additive exPlanations (SHAP) to analyze the impacts of variables. Further references can be consulted in~\cite{matthews2013risk},~\cite{yu2021measuring},~\cite{xie2022efficiency}, and~\cite{li2024evaluation}. 

\begin{figure}[H]
    \centering
    \includegraphics[width=1\textwidth]{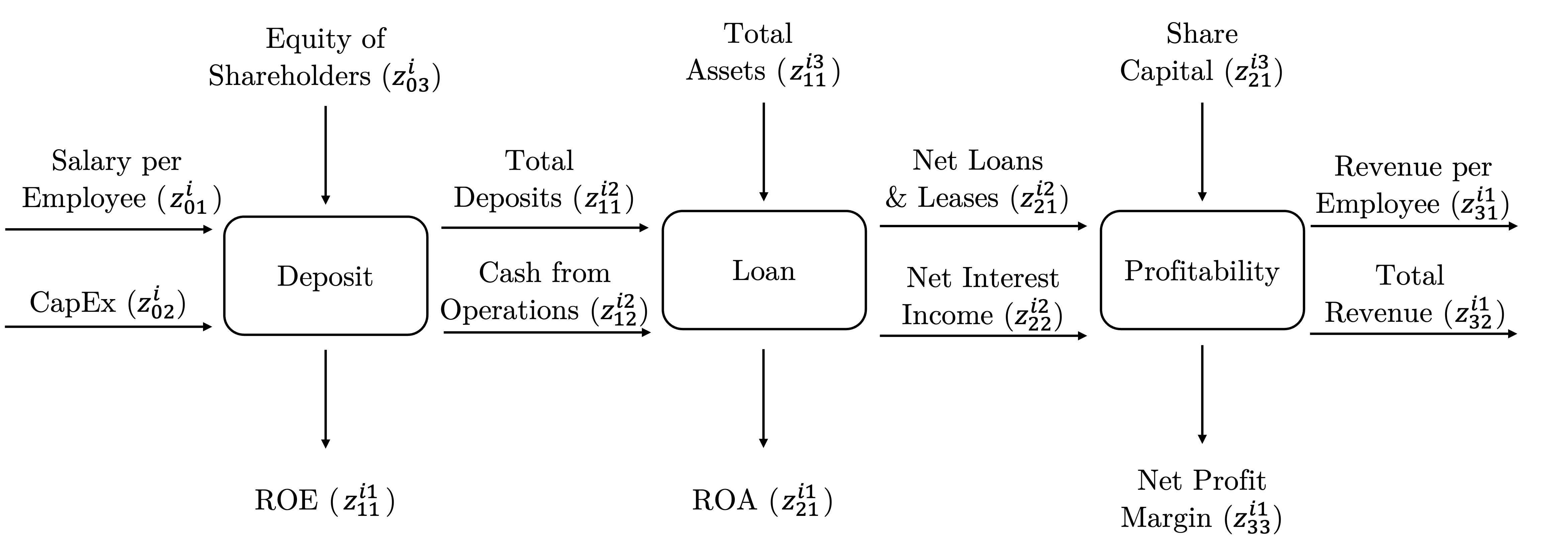} 
    \caption{Three-stage network DEA structure for FS}
    \label{fig:dea_structure}
\end{figure}

Bank operations encompass a diverse range of operations, and FinTech has carried distinct degrees of importance across these activities in recent years. As a result, FS has also been impacted by FinTech. In particular, the emergence of third-party payment platforms has simultaneously disrupted the ability of banks to attract deposits and challenged their profitability. These changes underscore the need for a comprehensive framework to evaluate FS across the stages of bank operations. For this purpose, we adopt the three-stage network DEA model by \cite{shi2025bank}. This approach decomposes the banking process into three stages: the deposit stage, the loan stage, and the profitability stage, based on the production approach~\citep{sealey1977inputs} and the intermediation approach~\citep{benston1965branch}. Figure~\ref{fig:dea_structure} illustrates the structure of this three-stage DEA model.

Specifically, following Figure~\ref{fig:dea_structure}, in the deposit stage, the focus is on capturing the FS factors involved in deposit-taking. This stage evaluates the ability of the bank to attract customer deposits and its capacity to manage deposits effectively. Salary per employee, capital expenditures (CapEx), and equity of shareholders are initial inputs, representing labor cost structure, capital investment, and financial strength. The outputs from this stage include total deposits, cash from operations, and return on equity (ROE). Subsequently, in the loan stage, the banks utilize the outputs of the deposit stage (deposits and operating cash) as primary funding sources, with total assets added as an external input. The outputs from this stage include net loan amount, net interest income, and ROA, reflecting credit issuance capacity and asset utilization efficiency. Finally, the profitability stage measures the FS of banks. It focuses on their ability to convert income from various sources into profits that benefits employees, customers, and investors. Given that most commercial banks in China are not publicly listed, we replace earnings per share with net profit margin and shares outstanding with share capital in the original model of \cite{shi2025bank}. Therefore, revenue per employee, total revenue, and net profit margin are treated as outputs that measure this ability.

We employ the additive efficiency decomposition method~\citep{cook2010network}, implemented for a three-stage network DEA~\citep{shi2025bank}, to model the structure in Figure~\ref{fig:dea_structure}. Let \( I = \{1, 2, \dots, |I|\} \) denote the set of decision-making units (DMUs), where each \( i \in I \) represents a bank, denoted as \( DMU_i \). Let \( P = \{1, \dots, |P|\} \) represent the set of stages in the process, with \( p \in P \) indexing each stage.
Following the network DEA framework established by \cite{cook2010network}, we classify the flows into and out of each stage $p$ as distinct vectors to ensure clear definition:

\begin{enumerate}[label=(\arabic*)]
    \item $R_p$-dimensional vector $Z_0$: The inputs enter the first stage ($p=1$);

    \item $R_p$-dimensional vector $Z_p^1$: The outputs are generated at stage $p$ and not passed to the stage $p+1$;

    \item  $S_p$-dimensional vector $Z_p^2$: The outputs are generated at stage $p$ and transferred as inputs to the stage $p+1$;

    \item $J_p$-dimensional vector $Z_p^3$: External inputs enter the process at the beginning of stage $p+1$. 
\end{enumerate}

The flow of these input and output vectors through the three-stage banking process is visually represented in Figure~\ref{fig:dea_structure}. The specific components of these vectors for $DMU_i$ are indexed as follows:

\begin{enumerate}[label=(\arabic*)]
    \item \( z_{pr}^{i1} \) denotes the \( r \)-th component (\( r = 1, \dots, R_p \)) of the \( R_p \)-dimensional output vector for \( DMU_i \) flowing from stage \( p \), which leaves the process at that stage \( p \), and is not passed on as an input to stage \( p+1 \). In the last stage \(|P| \), all the outputs are viewed as \( z_{|P|r}^{i1} \), as they leave the process;

    \item \( z_{pk}^{i2} \) denotes the \( k \)-th component (\( k = 1, \dots, S_p \)) of the \( S_p \)-dimensional output vector for \( DMU_i \) flowing from stage \( p \), and is passed on as a portion of inputs to stage \( p+1 \);

    \item \( z_{pj}^{i3} \) denotes the \( j \)-th component (\( j = 1, \dots, J_p \)) of the \( J_p \)-dimensional input vector for \( DMU_i \) at stage \( p+1 \), that enters the process at the beginning of that stage.
\end{enumerate}

Specifically, for any stage $p$ ($p \ge 2$), the total inputs are derived from the intermediate output $Z_{p-1}^{2}$ and the external input $Z_{p-1}^{3}$.
The weights for the above factors are defined as:

\begin{enumerate}[label=(\arabic*)]
    \item \( u_{pr} \) is the weight assigned to the output component \( z_{pr}^{i1} \) flowing from stage \( p \);

    \item \( \eta_{pk} \) is the weight for the output component \( z_{pk}^{i2} \) at stage \( p \), and it is also assigned to the same component which becomes an input to stage \( p+1 \);

    \item \( \nu_{pj} \) is the weight assigned to the \( j \)-th input component \( z_{pj}^{i3} \) that enters the process at the beginning of stage \( p+1 \).
\end{enumerate}

\begin{table}[H]
\centering
\caption{Notations in the three-stage network DEA-Malmquist model}
\label{tab:variables_explanation}
\renewcommand{\arraystretch}{1.2}
\begin{tabularx}{\textwidth}{@{}>{\centering\arraybackslash}p{3.2cm}X@{}}
\toprule
\textbf{Notation} & \textbf{Description} \\
\midrule
$i \in \{1, \ldots, |I|\}$ & Index for DMUs. \\ 
$p \in \{1, \ldots, |P|\}$ & Index for stages. \\
$r \in \{1, \ldots, R_p\}$ & Index for final output variables, where $R_p$ is the dimension of $Z_p^1$. \\ 
$k \in \{1, \ldots, S_p\}$ & Index for intermediate output variables, where $S_p$ is the dimension of $Z_p^2$. \\ 
$j \in \{1, \ldots, J_p\}$ & Index for external input variables, where $J_p$ is the dimension of $Z_p^3$. \\
$t \in \{1, \ldots, |T|\}$ & Index for years in the MI model. \\
\smash{$Z_0$} & The initial input vector to the first stage ($p=1$). \\
\smash{$Z_p^1$} & The $R_p$-dimensional output vector generated at stage $p$. \\
\smash{$Z_p^2$} & The $S_p$-dimensional output vector generated at stage $p$ that links to stage $p+1$. \\
\smash{$Z_p^3$} & The $J_p$-dimensional input vector that enters the process at the beginning of stage $p+1$. \\
\smash{$z_{pr}^{i1}$} & The $r$-th component of the final output vector $Z_p^1$ for $DMU_i$, flowing from stage $p$ and exiting the process. \\
\smash{$z_{pk}^{i2}$} & The $k$-th component of the output vector $Z_p^2$ for $DMU_i$, flowing from stage $p$ and passed on as an input to stage $p+1$. \\
\smash{$z_{pj}^{i3}$} & The $j$-th component of the external input vector $Z_p^3$ for $DMU_i$, which is an input to stage $p+1$ and enters the process at its beginning. \\
\smash{$u_{pr}$} & Weight for the $r$-th output $z_{pr}^{i1}$ at stage $p$. \\
\smash{$\eta_{pk}$} & Weight for the $k$-th output $z_{pk}^{i2}$ at stage $p$. \\
\smash{$\nu_{pj}$} & Weight for the $j$-th external input $z_{pj}^{i3}$ entering stage $p+1$. \\
\smash{$\theta_p$} & DEA estimated efficiency score of a $DMU_i$ at stage $p$. \\
\smash{$\theta$} & Weighted aggregate efficiency score across all stages, typically calculated using stage weights $w_1, w_2, w_3$. \\
\smash{$w_1, w_2, w_3$} & Weights assigned to the efficiency scores of stages 1, 2, and 3, respectively, where $w_1 + w_2 + w_3 = 1$. \\
\(\theta^{t}_{t+1}\) & Efficiency of the $DMU_i$ at time $t+1$, measured against the period $t$ technology frontier; captures intertemporal efficiency in the MI.\\
\(\theta_p^{t}(t+1)\) & Efficiency of the $DMU_i$ at time $t+1$ in stage $p$, measured against the period $t$ technology frontier; captures intertemporal efficiency in the MI. \\
\bottomrule
\end{tabularx}
\end{table}

When \(p = 2, 3, \dots\), the efficiency for \( DMU_i \) would be expressed as:

\begin{equation}
\theta_{p}=\frac{\sum_{r=1}^{R_{p}} u_{p r} z_{p r}^{i 1}+\sum_{k=1}^{S_{p}} \eta_{p k} z_{p k}^{i 2}+\varepsilon_{p}}{\sum_{k=1}^{S_{p-1}} \eta_{p-1 k} z_{p-1 k}^{i 2}+\sum_{j=1}^{J_{p}} \nu_{p-1 j} z_{p-1 j}^{i 3}},
\quad \varepsilon_{p} \text{ is unrestricted in sign}.
\label{eq:einstein1}
\end{equation}

The DEA formulations for each sub-stage corresponding to Equation~\eqref{eq:einstein1} are presented below:
\begin{subequations}
\begin{align}
\theta_1 &= 
\frac{u_{11} z_{11}^{i1} + \left( \eta_{11} z_{11}^{i2} + \eta_{12} z_{12}^{i2} \right) + \varepsilon_1}
{\nu_{01} z_{01}^{i} + \nu_{02} z_{02}^{i} + \nu_{03} z_{03}^{i}}, 
\quad \varepsilon_1 \text{ is unrestricted in sign} 
\label{eq:einstein2}, \\
\theta_2 &= 
\frac{u_{21} z_{21}^{i1} + \left( \eta_{21} z_{21}^{i2} + \eta_{22} z_{22}^{i2} \right) + \varepsilon_2}
{\left( \eta_{11} z_{11}^{i2} + \eta_{12} z_{12}^{i2} \right) + \nu_{11} z_{11}^{i3}},
\quad \varepsilon_2 \text{ is unrestricted in sign} 
\label{eq:einstein3}, \\
\theta_3 &= 
\frac{u_{31} z_{31}^{i1} + u_{32} z_{32}^{i1} + u_{33} z_{33}^{i1} + \varepsilon_3}
{\left( \eta_{21} z_{21}^{i2} + \eta_{22} z_{22}^{i2} \right) + \nu_{21} z_{21}^{i3}},
\quad \varepsilon_3 \text{ is unrestricted in sign},
\label{eq:einstein4}
\end{align}
\end{subequations}
where \(z_{0j}^{i}\) are the only inputs to first stage.

The overall performance $theta$ is computed as a weighted sum of the stage-specific efficiency scores, subject to a unit-sum constraint on the weights:
\begin{equation}
\theta = w_1 \theta_1 + w_2 \theta_2 + w_3 \theta_3, \quad
w_1 + w_2 + w_3 = 1.
\label{eq:overall_efficiency}
\end{equation}

We formulate a linear programming model to evaluate the overall efficiency across the three stages. Equation~\eqref{eq:vrs_model} maximizes the weighted sum of outputs and intermediate products across all stages. Equation~\eqref{eq:normalization} imposes a normalization condition, and equations~\eqref{eq:stage1_constraint}–\eqref{eq:stage3_constraint} give stage-specific feasibility constraints.
Equations~\eqref{eq:positive1_constraints}--\eqref{eq:epsilon_unrestricted} are domain constraints

\begin{subequations}
\begin{align}
\label{eq:vrs_model}
\max \quad & \sum_{p=1}^{P} \Big( \sum_{r=1}^{R_{p}} u_{pr} z_{pr}^{o1} 
+ \sum_{k=1}^{S_{p}} \eta_{pk} z_{pk}^{o2} + \varepsilon_p \Big) \\[1ex]
\text{s.t.} \quad
& \nu_{01} z_{01}^{o} + \nu_{02} z_{02}^{o} + \nu_{03} z_{03}^{o} 
  + \eta_{11} z_{11}^{o2} + \eta_{12} z_{12}^{o2} + \nu_{11} z_{11}^{o3} 
  + \eta_{21} z_{21}^{o2} + \eta_{22} z_{22}^{o2} + \nu_{21} z_{21}^{o3} = 1 \label{eq:normalization} \\[1ex]
& u_{11} z_{11}^{i1} + \eta_{11} z_{11}^{i2} + \eta_{12} z_{12}^{i2} + \varepsilon_1
  \le \nu_{01} z_{01}^{i} + \nu_{02} z_{02}^{i} + \nu_{03} z_{03}^{i}, \quad \forall i \in I \label{eq:stage1_constraint} \\
& u_{21} z_{21}^{i1} + \eta_{21} z_{21}^{i2} + \eta_{22} z_{22}^{i2} + \varepsilon_2 
  \le \eta_{11} z_{11}^{i2} + \eta_{12} z_{12}^{i2} + \nu_{11} z_{11}^{i3}, \quad \forall i \in I \label{eq:stage2_constraint} \\
& u_{31} z_{31}^{i1} + u_{32} z_{32}^{i1} + u_{33} z_{33}^{i1} + \varepsilon_3
  \le \eta_{21} z_{21}^{i2} + \eta_{22} z_{22}^{i2} + \nu_{21} z_{21}^{i3}, \quad \forall i \in I \label{eq:stage3_constraint} \\[1ex]
& u_{11}, u_{21}, u_{31}, u_{32}, u_{33} > 0  
\label{eq:positive1_constraints}\\
& \eta_{11}, \eta_{12}, \eta_{21}, \eta_{22} > 0  \label{eq:positive2_constraints}\\
& \nu_{01}, \nu_{02}, \nu_{03}, \nu_{11}, \nu_{21} > 0  \label{eq:positive3_constraints} \\
&\varepsilon_1, \varepsilon_2, \varepsilon_3 \text{ are unrestricted in sign} \label{eq:epsilon_unrestricted}
\end{align}
\end{subequations}

Traditional DEA models suffer from the inherent limitation of being static and failing to capture intertemporal performance dynamics~\citep{fare1994productivity}. Thus, we incorporate the MI, allowing us to track how bank performance evolves over time. The MI is widely applied to measure the changes in productivity of banks~\citep{caves1982economic}. Studies include analyses of credit banks in Japan~\citep{barros2009productivity}, and \cite{bansal2022dynamic} use a dynamic network DEA-based Malmquist–Luenberger index to measure the productivity changes of Indian banks. The MI measures efficiency changes between two time periods by calculating the ratio of the distances of each data point to a common technology frontier~\citep{casu2004productivity}. The overall MI is defined for periods \(t \in T\), where \(T = \{1, 2, \dots, |T|-1\}\), as follows:
\begin{equation}
MI_{t,t+1} = \sqrt{
\frac{\theta^{t}_{t+1}}{\theta^{t}_{t}} \cdot
\frac{\theta^{t+1}_{t+1}}{\theta^{t+1}_{t}}
}.
\label{eq:mpi_overall}
\end{equation}
Let \(\theta^{t}_{t+1}\) denote the efficiency of \(DMU_i\) at period \(t+1\) evaluated under the technology of period \(t\), and the other \(\theta\) terms in the formula are interpreted similarly.

We use MI to analyze the changes in FS of Chinese commercial banks. Changes in FS can be assessed by an output-oriented or input-oriented approach. The former way measures how the actual output compares to the maximum possible output achievable with the same inputs and technology. On the contrary, the latter one measures productivity changes when the same output is produced with fewer inputs under a given technology. According to~\cite{jaffry2007regulatory}, an output-oriented model is more suitable for developing countries. Therefore, we adopt an output-oriented approach. In the following, Equation~\eqref{eq:mpi_stage_p} for each production stage \( p \in P \) is used:
\begin{equation}
MI_{t,t+1}^{p} = \sqrt{
\frac{\theta_p^{t}(t+1)}{\theta_p^{t}(t)} \cdot
\frac{\theta_p^{t+1}(t+1)}{\theta_p^{t+1}(t)}
},
\label{eq:mpi_stage_p}
\end{equation}
where \(\theta_p^{t}(t+1)\) denotes the efficiency of \(DMU_i\) at stage \(p\) and period \(t+1\), evaluated under the technology of period \(t\), and the other \(\theta\) terms in the formula are interpreted similarly.

Accordingly, an MI value greater than one indicates a positive trend in FS, a value equal to one indicates no change, and a value less than one indicates a decline relative to the prior period. 
The overall measure of FS can be decomposed into technical change (TC) and efficiency change (EC). 
The TC reflects improvements in technology and shifts in the production frontier, and the EC captures the catching-up effect, indicating whether banks move closer to or farther from the best-practice frontier. TC and EC are computed as follows:
\begin{equation}
\label{eq:mpi_decomp}
\displaystyle
MI_{t,t+1} = 
\underbrace{\frac{\theta^{t+1}_{t+1}}{\theta^{t}_{t}} \vphantom{\sqrt{\frac{\theta^{t}_{t+1}}{\theta^{t+1}_{t+1}} \cdot \frac{\theta^{t}_{t}}{\theta^{t+1}_{t}}}}}_{\text{Efficiency Change (EC)}} 
\;\times\;
\underbrace{\sqrt{
\frac{\theta^{t}_{t+1}}{\theta^{t+1}_{t+1}} \cdot
\frac{\theta^{t}_{t}}{\theta^{t+1}_{t}}
}}_{\text{Technical Change (TC)}}.
\end{equation}

In empirical studies of the banking sector, researchers often employ TC and EC. For instance, \cite{portela2010malmquist} evaluate productivity changes in Portuguese bank branches across different periods and branches by TC and EC. \cite{assaf2013turkish} compare TC and EC across different types of banks in Turkey. Given the above studies, we also adopt this approach as part of our robustness tests in Section~\ref{subsec:Robustness}.

\subsection{Two-way fixed effects model}
To examine the relationship between FS and FinTech, we run a two-way fixed effects regression using panel data. Reviewing existing literature~\citep{baltagi2008econometric, wooldridge2010econometric}, we find that unobserved heterogeneity across units and periods may bias estimation results. The two-way fixed effects model mitigates this issue by controlling for both individual and time effects. Let \(i\) and \(t\) denote the evaluated commercial bank and year, respectively. 
The dependent variable is $FSI$ (FS index), which represents the FS of commercial banks. To this end, the regression model is specified as follows:

\begin{equation}
FSI_{i,t} = \beta_0 + \beta_1 FTI_{i,t}
+ \sum_{c=2}^{3} \beta_c \, M_{c,t} + \sum_{c=4}^{10} \beta_c \, X_{c,i,t} 
+ \delta_i + \mu_t + e_{i,t},  \quad \forall i \in I,\forall t \in T.
\label{eq:baseline_model}
\end{equation}

In Equation~\eqref{eq:baseline_model}, the explanatory variable is \(FTI\) (FinTech index). The set of control variables consists of two parts: macroeconomic variables and bank-level variables. The macroeconomic variables are represented by \(M_{c,t}\), and the bank-level variables by \(X_{c,i,t}\). The subscript \(c\) indexes different control variables within each category, while \(e_{i,t}\) denotes the stochastic error term. Moreover, \(\delta_i\) and \(\mu_t\) denote the individual (bank-specific) and time fixed effects, respectively. The sign and significance of \(\beta_1\) are used to examine the relationship between FinTech and the FS of commercial banks, 
and $\beta_0$ is the constant term.
Standard errors are clustered at the individual level to address within-group correlation and to avoid underestimating standard errors.

We primarily evaluate the FS of commercial banks and analyze the impact of FinTech thereon. The dependent variable $FSI$ is derived from the three-stage network DEA-Malmquist model. We use the Peking University Digital Financial Inclusion Index compiled by \cite{guo2020measuring} to represent FinTech~\citep{lee2023fintech, hu2024will}. Notably, this index reflects the extent of digital delivery and accessibility of financial services across regions, aligning with the core dimensions of technology-driven financial development. We standardize the index by dividing its original values by 100 to control for scale differences and ensure regional comparability. The resulting standardized index is the core explanatory variable ($FTI$).

The empirical analysis and robustness checks incorporate a set of control variables consistent with prior studies~\citep{cheng2020does,wang2021can,lee2023fintech}. At the macroeconomic level, the prefecture-level GDP growth rate ($GDP_g$) and the financial development level ($FDL$) are included. Since banks operate in different regions with varying economic and capital market conditions, we include these variables as controls to mitigate potential bias caused by regional differences. At the individual bank level, variables such as loan-to-deposit ratio ($LDR$), non-interest income ratio ($NIIR$), return on assets ($ROA$), debt-to-asset ratio ($DAR$), total assets ($TAS$), operating expenses ($OEX$), and capital adequacy ratio ($CAR$) are included. These variables account for differences in profitability, bank size, risk management, and other related aspects. Controlling these factors reduces bias due to individual bank differences. Definitions and calculation methods for these variables are presented in Table~\ref{tab:variable_definitions_en}.

\begin{table}[H]
\centering
\small
\caption{Variables of the two-way fixed effects model}
\label{tab:variable_definitions_en}
\renewcommand{\arraystretch}{1.2}
\begin{tabularx}{\textwidth}{lllX} 
\toprule
\textbf{Category} & \textbf{Variable Name} & \textbf{Abbreviation} & \textbf{Definition} \\
\midrule
Dependent Variable & Financial Sustainability Index & $FSI$ & Composite financial indicators calculated by a three-stage network DEA-Malmquist model \\
\addlinespace
Explanatory Variable & FinTech Index & $FTI$ & The Peking University Digital Financial Inclusion Index divided by 100 \\
\addlinespace
Control Variables
    & GDP Growth Rate (the prefecture-level) & $GDP_g$ & (Current GDP - Previous GDP) / Previous GDP \\
    & Financial Development Level & $FDL$ & Ratio of total deposits and loans to local GDP \\
    & Loan-to-Deposit Ratio & $LDR$ & Total loans / Total deposits \\
    & Non-interest Income Ratio & $NIIR$ & Non-interest income / Operating income \\
    & Return on Assets & $ROA$ & Net profit / Total assets \\
    & Debt-to-Asset Ratio & $DAR$ & Total liabilities / Total assets \\
    & Total Assets & $TAS$ & Natural logarithm of total assets at year-end \\
    & Operating Expenses & $OEX$ & Natural logarithm of operating expenses at year-end \\
    & Capital Adequacy Ratio & $CAR$ & Eligible capital / Risk-weighted assets \\
\bottomrule
\end{tabularx}
\end{table}

\section{Empirical analysis}
\label{sec:analysis}
This section presents an empirical analysis to identify the impact, underlying mechanisms, and heterogeneous effects of FinTech on FS. Firstly, the values of FS are computed based on the methodology in Section~\ref{sec:data}. We also conduct descriptive statistics for relevant variables and run a fixed-effects panel regression to examine the impact of FinTech development. Secondly, a series of robustness checks is performed to verify the reliability of the baseline regression results. Thirdly, stage-specific MIs are introduced as channel variables to investigate the transmission mechanisms, where FinTech affects bank financial performance across different operational stages. Finally, heterogeneity analyses are conducted based on bank listing status and patent ownership to explore differential impacts across bank types.

\subsection{Main results}
\label{subsec:influence}
\textbf{FS estimation results:} Table~\ref{tab:mpi_summary} reports the annual averages of FS for 104 banks from 2015 to 2023. The mean MI exceeds one in most years, indicating an overall upward trend in FS of the banking sector. The highest MI value is 1.3287 in 2020, suggesting that the financial capability of banks demonstrated strong resilience during the initial period of the COVID-19 pandemic, in line with evidence of policy-supported stability in Chinese banks~\citep{wu2020effect}. In contrast, the MI falls sharply to 0.7530 in 2021, the lowest level during the sample period. This deterioration may be associated with pressures on FS in the post-pandemic period. These pressures may stem from heightened asset quality risks, tighter regulatory oversight, and shifts in credit allocation patterns, which are supported by the findings of \citet{elnahass2021global} and \citet{yao2025impact}. By 2023, the MI returns to 1.0016, a near-neutral level, showing that FS in the banking sector has little change compared to the previous year.

\begin{table}[H]
\centering
\caption{Annual average MI for 104 banks from 2015 to 2023}
\label{tab:mpi_summary}
\renewcommand{\arraystretch}{1.2}
\begin{tabularx}{\textwidth}{*{4}{>{\centering\arraybackslash}X}}  
\toprule
\textbf{Year} & \textbf{MI} & \textbf{TC} & \textbf{EC} \\
\midrule
2015 & 1.1311 & 1.0847 & 1.0572 \\
2016 & 0.8056 & 0.8242 & 0.9567 \\
2017 & 1.1678 & 1.1693 & 1.0109 \\
2018 & 1.0798 & 1.0561 & 1.0251 \\
2019 & 1.0760 & 1.0872 & 0.9973 \\
2020 & 1.3287 & 1.3368 & 1.0066 \\
2021 & 0.7530 & 0.6963 & 1.0930 \\
2022 & 1.0677 & 1.0531 & 1.0183 \\
2023 & 1.0016 & 1.0250 & 0.9867 \\
\bottomrule
\multicolumn{4}{p{\textwidth}}{\footnotesize 
\textbf{Notes:} This table reports the annual average values of the MI and its two components: TC and EC, over the period 2015–2023. An MI value greater than one indicates an improvement in FS performance relative to the previous year, while a value less than one indicates a decline.
}
\end{tabularx}
\end{table}

\noindent\textbf{Descriptive statistical analysis:} The descriptive statistics of the variables are reported in Table~\ref{tab:desc_stats}. Specifically, the mean of $FSI$ is 1.0457, with a standard deviation of 0.2781, indicating that the FS of Chinese commercial banks in the sample remains relatively stable. For $FTI$, the average value is 2.7769 with a standard deviation of 0.4801, suggesting some variation in the development of FinTech across regions. Regarding the control variables, $NIIR$ displays considerable variation across banks, while $CAR$ and $DAR$ are relatively stable, likely reflecting regulatory consistency. It is worth noting that the minimum value of $GDP_g$ is \(-5.6\), which occurred in Shenyang in 2016, highlighting the existence of substantial regional economic disparities and underscoring the necessity of controlling for such heterogeneity in subsequent analyses. Given the focus of this study, i.e., examining how FinTech affects the FS of commercial banks, the detailed statistical characteristics of control variables are not further elaborated here.

\begin{table}[H]
\centering
\caption{Descriptive statistics of all variables}
\label{tab:desc_stats}
\small
\renewcommand{\arraystretch}{1.2} 
\begin{tabularx}{\textwidth}{*{6}{>{\centering\arraybackslash}X}} 
\toprule
\textbf{Variable} & \textbf{Sample size} & \textbf{Mean} & \textbf{Std. Dev.} & \textbf{Min} & \textbf{Max} \\
\midrule
$FSI$      & 936 & 1.0457  & 0.2781 & 0.5464  & 1.6356 \\
\addlinespace
$FTI$ & 936 & 2.7769  & 0.4801 & 1.5221  & 3.7322 \\
\addlinespace
$GDP_g$   & 935 & 6.2198  & 2.5265 & -5.6000 & 12.5000 \\
\addlinespace
$FDL$      & 927 & 4.2641  & 1.5934 & 1.4265  & 7.9760 \\
\addlinespace
$LDR$     & 935 & 0.7492  & 0.1215 & 0.5263  & 0.9809 \\
\addlinespace
$NIIR$    & 936 & 21.3664 & 12.5907 & 3.6860 & 51.5334 \\
\addlinespace
$ROA$     & 936 & 0.6973  & 0.2270 & 0.2532  & 1.0747 \\
\addlinespace
$DAR$     & 936 & 0.9255  & 0.0135 & 0.7824  & 0.9629 \\
\addlinespace
$TAS$     & 936 & 26.8395 & 1.5258 & 24.0439 & 31.4309 \\
\addlinespace
$OEX$     & 936 & 22.6281 & 1.5297 & 19.6937 & 26.9756 \\
\addlinespace
$CAR$     & 936 & 13.4558 & 1.9846 & 2.3700  & 33.8600 \\
\addlinespace
\bottomrule
\multicolumn{6}{p{\textwidth}}{\footnotesize 
\textbf{Notes:} This table shows the explanation and descriptive statistics of all variables. Differences in sample size are due to missing values in control variables. 
}
\end{tabularx}
\end{table}

\noindent\textbf{Baseline regression:} According to Equation \eqref{eq:baseline_model}, we use panel data to test the relation between FinTech and FS, and Table~\ref{tab:regression_with_se} presents results of a two-way fixed effects model. Columns (1) and (2) report results without additional control variables, whereas Columns (3) and (4) include them. Moreover, we cluster standard errors at the bank level in all estimations. Columns (1) and (3) do not include individual fixed effects, while Columns (2) and (4) control for both individual and time fixed effects.

Several important observations can be drawn. First, the coefficients of $FTI$ on $FSI$ are negative and statistically significant across all specifications, regardless of whether control variables or fixed effects are included. This relationship remains stable even after accounting for potential bank-level and time-varying confounders. Further, in Column (4), which incorporates individual and time fixed effects as well as additional controls, the coefficient on $FTI$ is -0.540, and statistically significant at the 1\% level. This result implies that a one-unit increase in FinTech development is associated with a 0.54-point decline in $FSI$, i.e., a non-trivial magnitude relative to its standard deviation. Economically, FinTech may weaken FS of commercial banks by intensifying disintermediation and shifting customers toward technology-based financial services. This potential mechanism will be empirically examined in the following analysis. Finally, among the control variables, $ROA$ and $OEX$ are positively and significantly associated with $FSI$, confirming the importance of profitability and operational investment for maintaining stability. By contrast, $TAS$ exhibits a significantly negative coefficient, suggesting that excessive asset expansion can undermine financial stability through resource misallocation and heightened financial risk. The primary finding remains robust, as the negative impact of $FTI$ persists across all model specifications.

\begin{table}[H]
\centering
\caption{Baseline regression}
\label{tab:regression_with_se}
\small
\begin{tabularx}{\textwidth}{l *{4}{>{\centering\arraybackslash}X}}
\toprule
            & (1) & (2) & (3) & (4) \\
\cmidrule(lr){2-3} \cmidrule(lr){4-5}
            & \multicolumn{2}{c}{\textbf{FinTech Only}} & \multicolumn{2}{c}{\textbf{Full Controls}} \\
            & \multicolumn{2}{c}{$FSI$} & \multicolumn{2}{c}{$FSI$} \\
\midrule
$FTI$   & -0.042\sym{**} & -0.530\sym{***} & -0.062\sym{***} & -0.540\sym{***} \\
            & (0.017)        & (0.189)         & (0.023)         & (0.203)         \\
[1em]
$GDP_g$     &                &                 & -0.036\sym{***} & 0.005           \\
            &                &                 & (0.004)         & (0.007)         \\
[1em]
$FDL$        &                &                 & -0.002          & -0.004          \\
            &                &                 & (0.006)         & (0.028)         \\
[1em]
$LDR$       &                &                 & 0.063           & 0.167           \\
            &                &                 & (0.083)         & (0.129)         \\
[1em]
$NIIR$      &                &                 & -0.000          & -0.001          \\
            &                &                 & (0.001)         & (0.001)         \\
[1em]
$ROA$       &                &                 & 0.184\sym{***}  & 0.166\sym{**}   \\
            &                &                 & (0.050)         & (0.075)         \\
[1em]
$DAR$       &                &                 & 1.579           & 2.783\sym{*}    \\
            &                &                 & (0.982)         & (1.562)         \\
[1em]
$TAS$       &                &                 & -0.151\sym{***} & -0.254\sym{**}  \\
            &                &                 & (0.039)         & (0.108)         \\
[1em]
$OEX$       &                &                 & 0.130\sym{***}  & 0.196\sym{***}  \\
            &                &                 & (0.036)         & (0.063)         \\
[1em]
$CAR$       &                &                 & 0.012\sym{**}   & 0.010           \\
            &                &                 & (0.006)         & (0.007)         \\
[1em]
Bank FE     & NO             & YES             & NO              & YES             \\
[1em]
Year FE     & YES            & YES             & NO              & YES             \\
[1em]
Constant term      & 1.163\sym{***} & 2.517\sym{***}  & 0.766           & 1.996           \\
            & (0.048)        & (0.525)         & (0.866)         & (2.471)         \\
\midrule
Observations         & 936            & 936             & 925             & 925             \\
$R^{2}$     & 0.06           & 0.42            & 0.10            & 0.45            \\
\bottomrule
\multicolumn{5}{p{\textwidth}}{\footnotesize 
\textbf{Notes:} The dependent variable is the $FSI$ rating from the three-stage network DEA-Malmquist model. Columns (1) and (2) report results without control variables, while Columns (3) and (4) include control variables. Differences in sample size are due to missing values in control variables. The main model controls for individual and time fixed effects. Variable definitions are provided in Table~\ref{tab:variable_definitions_en}. \sym{*}, \sym{**}, and \sym{***} indicate statistical significance at the 10\%, 5\%, and 1\% levels, respectively. Standard errors are clustered at the bank level and reported in parentheses. $R^2$ represents the coefficient of determination.
}
\end{tabularx}
\end{table}

\subsection{Robustness tests}
\label{subsec:Robustness}
We perform a series of robustness tests to validate the consistency of the main findings. Specifically, we examine whether the results remain robust when using alternative TC and EC measures from Equation~\eqref{eq:mpi_decomp}. These measures are components decomposed from the MI applied to $FSI$. Furthermore, we alleviate potential endogeneity issues by implementing an instrumental variable (IV) approach.

\noindent\textbf{IV approach:} We include a battery of control variables and apply a two-way fixed effects model, but the estimation is still influenced by unobserved heterogeneity or omitted variables. To mitigate these problems, we adopt a strategy with instrumental variables. Specifically, we implement a two-stage least squares (2SLS) method~\citep{wooldridge2010econometric} and a control function (CF) method~\citep{wooldridge2015control} to obtain more consistent estimates.

We introduce two instrumental variables. The first instrumental variable ($IV_1$) is the one-period lag of $FTI$. Due to the temporal precedence over bank financial outcomes, $IV_1$ helps alleviate potential endogeneity bias. The second instrument ($IV_2$) is the logarithm of the interaction between the distance to Hangzhou, a recognized FinTech hub, and the average annual Digital Financial Inclusion Index (excluding the city itself). This variable captures the regional spillover effects of FinTech development. To ensure the reliability of the instrumental variables estimation, several diagnostic tests are conducted. The Kleibergen–Paap rk LM and rk Wald F statistics are used to assess underidentification and weak instrument issues, respectively, while the Hansen J test examines instrument validity. As shown in the Column (1) of Table~\ref{tab:Robustness_result}, both instruments strongly correlate with $FTI$. The Kleibergen-Paap rk LM statistic rejects the null of underidentification ($p$ \textless 0.01), and the rk Wald F statistic (27.377) exceeds the Stock-Yogo critical value of 19.93, suggesting no weak instrument issue. The Hansen J test also yields an insignificant $p$-value (0.1906), confirming the validity of the instruments. Then, let $\widehat{FTI}_{i,t}$ denote the predicted value from the first-stage regression, and $\delta_i$ and $\mu_t$ indicate the fixed effects of individual and time, respectively. The instrumental variables include $IV_1$ and $IV_2$. All other control variables are consistent with those used in the baseline regressions. The IV approach is conducted by estimating the following 2SLS method Equations~\eqref{eq:first_stage} and~\eqref{eq:second_stage}:

\begin{align}
&FTI_{i,t} = \rho_0 + \rho_1 \, IV_{1,i,t} + \rho_2 \, IV_{2,i,t} 
+ \sum_{c=1}^{2} \beta_c \, M_{c,t} + \sum_{c=3}^{9} \beta_c \, X_{c,i,t}+ \delta_i + \mu_t + \xi_{i,t}, &&\forall i \in I,\forall t \in T,
\label{eq:first_stage}
\\
&FSI_{i,t} = \beta_0 + \beta_1 \, \widehat{FTI}_{i,t} 
+ \sum_{c=2}^{3} \beta_c \, M_{c,t} + \sum_{c=4}^{10} \beta_c \, X_{c,i,t} + \delta_i + \mu_t + e_{i,t}, &&\forall i \in I,\forall t \in T.
\label{eq:second_stage}
\end{align}

As reported in the Column (2) of Table~\ref{tab:Robustness_result}, the estimated coefficient on $FTI$ remains significantly negative after controlling for endogeneity. Overall, the results of this analysis are consistent with the main findings, namely that FinTech negatively affects FS.

In addition to 2SLS method, we also adopt the CF method as a complementary strategy to alleviate potential endogeneity. Compared to 2SLS method, the CF method offers greater flexibility when dealing with heteroskedasticity or complex error structures. The CF method is specified in Equations~\eqref{eq:cf_first_stage}–\eqref{eq:cf_second_stage}. Let $\widehat{\xi}_{i,t}$ denote the residuals from the first-stage regression, capturing endogeneity effects, while the remaining variables are consistent with those in the baseline regressions, we obtain

\begin{align}
&FTI_{i,t} = \rho_0 + \rho_1 \, IV_{1,i,t} + \rho_2 \, IV_{2,i,t} 
+ \sum_{c=1}^{2} \beta_c \, M_{c,t} + \sum_{c=3}^{9} \beta_c \, X_{c,i,t} + \delta_i + \mu_t + \xi_{i,t}, &&\forall i \in I, \forall t \in T,
\label{eq:cf_first_stage} 
\\
&FSI_{i,t} = \beta_0 + \beta_1 \, FTI_{i,t} + \lambda \, \widehat{\xi}_{i,t} 
+ \sum_{c=2}^{3} \beta_c \, M_{c,t} + \sum_{c=4}^{10} \beta_c \, X_{c,i,t} + \delta_i + \mu_t + e_{i,t}, &&\forall i \in I, \forall t \in T.
\label{eq:cf_second_stage}
\end{align}

We present the results in the Column (3) of Table~\ref{tab:Robustness_result}. The coefficient on $FTI$ remains significantly negative, consistent with the baseline and 2SLS method estimates. Importantly, the residual term from the second stage is statistically significant, which suggests that the original $FTI$ variable suffers from endogeneity. This finding suggests that the CF method successfully corrects for this bias. Together, these findings reinforce the robustness of the adverse effect of FinTech development on the FS of commercial banks.

\noindent\textbf{Alternative measure of FS:} Given that it is calculated using a three-stage network DEA-Malmquist model, we examine the influence of the internal components of the MI. Specifically, we decompose the MI into two sub-indicators, TC and EC, and replace $FSI$ with them in the regression model for robustness checks. These two components reflect technological progress over time and changes in managerial efficiency. Table~\ref{tab:mpi_summary} presents the annual means of these two indices. As reported in Columns (4) and (5) of Table~\ref{tab:Robustness_result}, the signs of the key explanatory variables remain unchanged, despite the statistical significance of some coefficients declining compared to the baseline results. This result suggests that the main findings of this study, namely the negative effect of FinTech on FS of banks, are reasonably robust to alternative specifications.

\begin{table}[H]
\centering
\normalsize
\caption{Robustness tests}
\label{tab:Robustness_result}
\renewcommand{\arraystretch}{1.1}
\begin{tabularx}{\textwidth}{>{\raggedright\arraybackslash}X *{5}{>{\centering\arraybackslash}X}}
\toprule
            & \shortstack[c]{(1)\\2SLS \& CF 1st} 
            & \shortstack[c]{(2)\\2SLS 2nd} 
            & \shortstack[c]{(3)\\CF 2nd} 
            & \shortstack[c]{(4)\\TC} 
            & \shortstack[c]{(5)\\EC} \\
\midrule
$FTI$     &                & -0.878\sym{**}     & -0.878\sym{***}    & -0.342\sym{*}       & -0.237\sym{*}       \\
            &                & (0.405)            & (0.304)            & (0.181)             & (0.124)             \\
$IV_1$       & 0.559\sym{***} &                     &                     &                     &                     \\
            & (0.069)        &                     &                     &                     &                     \\
$IV_2$       & -0.117\sym{***}&                     &                     &                     &                     \\
            & (0.034)        &                     &                     &                     &                     \\
$\widehat{\xi}_{i,t}$ &           &                    & 0.840\sym{**}      &                     &                     \\
            &                &                    & (0.374)            &                     &                     \\
$GDP_g$      & 0.002\sym{*}   & 0.007              & 0.007              & -0.005              & 0.011\sym{**}       \\
            & (0.001)        & (0.007)            & (0.007)            & (0.004)             & (0.005)             \\
$FDL$        & -0.001         & -0.013             & -0.013             & 0.006               & -0.006              \\
            & (0.005)        & (0.028)            & (0.026)            & (0.017)             & (0.016)             \\
$LDR$       & -0.057         & 0.145              & 0.145              & -0.003              & 0.203\sym{**}       \\
            & (0.035)        & (0.165)            & (0.138)            & (0.111)             & (0.086)             \\
$NIIR$      & -0.000         & -0.002\sym{*}      & -0.002\sym{*}      & -0.001              & -0.000              \\
            & (0.000)        & (0.001)            & (0.001)            & (0.001)             & (0.001)             \\
$ROA$       & 0.033\sym{**}  & 0.304\sym{***}     & 0.304\sym{***}     & -0.081              & 0.205\sym{***}      \\
            & (0.013)        & (0.082)            & (0.072)            & (0.052)             & (0.053)             \\
$DAR$       & 0.238          & 2.780\sym{*}       & 2.780              & -0.114              & 2.822\sym{***}      \\
            & (0.167)        & (1.559)            & (1.705)            & (1.099)             & (0.868)             \\
$TAS$       & 0.004          & -0.289\sym{***}    & -0.289\sym{**}     & -0.069              & -0.225\sym{***}     \\
            & (0.017)        & (0.111)            & (0.116)            & (0.072)             & (0.059)             \\
$OEX$       & -0.006         & 0.220\sym{***}     & 0.220\sym{***}     & 0.050               & 0.144\sym{***}      \\
            & (0.009)        & (0.065)            & (0.067)            & (0.051)             & (0.042)             \\
$CAR$       & 0.002\sym{**}  & 0.007              & 0.007              & 0.005               & 0.003               \\
            & (0.001)        & (0.007)            & (0.007)            & (0.006)             & (0.005)             \\
\midrule
Observations & 821 & 821 & 821 & 925 & 925 \\
\(R^2\)      & 0.99 & 0.45 & 0.49 & 0.57 & 0.17 \\
\bottomrule
\end{tabularx}
\vspace{1ex}
\begin{tabularx}{\textwidth}{l *{5}{>{\centering\arraybackslash}X}}
\multicolumn{6}{l}{IV diagnostics for Columns (1)–(3):} \\
Kleibergen-Paap rk LM statistic       & \multicolumn{5}{c}{112.984 ($p$-value = 0.0000)} \\
Cragg-Donald Wald F statistic         & \multicolumn{5}{c}{205.381 (19.93)} \\
Kleibergen-Paap rk Wald F statistic   & \multicolumn{5}{c}{27.377 (19.93)} \\
Hansen J statistic                     & \multicolumn{5}{c}{1.713 ($p$-value = 0.1906)} \\
\bottomrule
\end{tabularx}
\vspace{0.5ex}
\parbox{\textwidth}{\footnotesize 
\textbf{Notes:} This table shows the results of robustness tests. 
Column (1) reports the first-stage regressions of 2SLS and the CF methods. 
The numbers in parentheses after the Cragg–Donald and Kleibergen–Paap rk Wald F statistics indicate 
the Stock–Yogo 10\% maximal instrumental variables size distortion critical value. 
Variable definitions are provided in Table~\ref{tab:variable_definitions_en}. 
$\widehat{\xi}_{i,t}$ denotes the residual term from the first stage of the CF method. 
Differences in sample size are due to missing values in control variables and using one-period lagged explanatory variables. 
All regressions include individual and time fixed effects. 
*, **, and *** indicate significance at the 10\%, 5\%, and 1\% levels, respectively. Standard errors are clustered at the bank level and reported in parentheses. $R^2$ represents the coefficient of determination.
}
\end{table}

\subsection{Economic mechanisms}
\label{subsec:Mechanism discussion}
Building on the baseline regression results, which indicate a negative relationship between FinTech and overall FS of banks, we further explore the potential transmission mechanisms. Following the approach of \cite{liang2017foundations}, we conduct separate regressions in two steps using the stage-level MIs, corresponding to the deposit ($MI_d$), loan ($MI_l$), and profitability ($MI_p$) stages. These indices serve as the channel through which FinTech influences FS of banks. In the first stage, each channel variable is regressed on $FTI$ to obtain the component explained by FinTech. In the second stage, we regress $FSI$ on the predicted values of channel variables from the first stage. These predicted values capture the part of FS variation that operates through FinTech-driven channels. Control variables are included in both stages. While this strategy resembles an IV approach in structure, FinTech is not formally used as an instrument for channel variables. The specific two-stage model is formally presented in Equations~\eqref{eq:first_stage1} and \eqref{eq:second_stage2}:

\begin{align}
&MI^{(\phi)}_{i,t} = \alpha_0^{(\phi)} + \alpha_1^{(\phi)} \, FTI_{i,t} 
+ \sum_{c=1}^{2} \beta_c \, M_{c,t} + \sum_{c=3}^{9} \beta_c \, X_{c,i,t}
+ \delta_i + \mu_t + e^{(\phi)}_{i,t},  &&\forall i \in I, \forall t \in T,
\label{eq:first_stage1} \\
&FSI_{i,t} = \kappa_0^{(\phi)} + \kappa_1^{(\phi)} \, \widehat{MI}^{(\phi)}_{i,t} 
+ \sum_{c=1}^{2} \beta_c \, M_{c,t} + \sum_{c=3}^{9} \beta_c \, X_{c,i,t}  
+ \delta_i + \mu_t + e_{i,t}^{(\phi,2)},&&\forall i \in I, \forall t \in T,
\label{eq:second_stage2}
\end{align}
\noindent where $\phi \in \{d, l, p\}$\footnotemark.

\footnotetext{%
The superscript $\phi \in {d, l, p}$ denotes the three sub-stages of the FS of banks: the deposit ($d$), the loan ($l$), and the profitability ($p$) stages. Each $e^{(\phi)}_{i,t}$ represents the regression residual corresponding to the respective stage.
}

The dynamic patterns of banking efficiency across different operational stages are illustrated in Table~\ref{tab:malmquist_stages}, which reports the annual averages of MIs for each stage from 2015 to 2023. As shown, the indices in most years exceed one across all stages, suggesting a generally positive performance in FS across stages. In 2020, $MI_d$ peaked at 1.7710, reflecting a sharp improvement in deposit efficiency during the early period of the COVID-19 pandemic. In contrast, $MI_p$ dropped to 0.7104 in 2021, which is the lowest among all years, indicating a substantial decline in the earning capacity of banks. This deterioration is likely attributable to mounting pressure from non-performing loans, narrowing interest margins, and a tightening regulatory environment.

\begin{table}[H]
\centering
\caption{Annual average MIs for each stage from 2015 to 2023}
\label{tab:malmquist_stages}
\renewcommand{\arraystretch}{1.2}
\begin{tabularx}{\textwidth}{*{4}{>{\centering\arraybackslash}X}}
\toprule
\textbf{Year} & \textbf{Deposit stage} & \textbf{Loan stage} & \textbf{Profitability stage} \\
             & \textbf{$MI_d$}        & \textbf{$MI_l$}     & \textbf{$MI_p$} \\
\midrule
2015 & 1.1972 & 1.1700 & 1.1643 \\
\addlinespace
2016 & 0.7285 & 0.8477 & 0.9187 \\
\addlinespace
2017 & 1.4727 & 1.0662 & 1.1391 \\
\addlinespace
2018 & 1.0590 & 1.0821 & 1.1605 \\
\addlinespace
2019 & 1.0331 & 1.0787 & 1.1783 \\
\addlinespace
2020 & 1.7710 & 1.1711 & 1.2709 \\
\addlinespace
2021 & 0.7473 & 0.8932 & 0.7104 \\
\addlinespace
2022 & 1.0021 & 1.0398 & 1.2527 \\
\addlinespace
2023 & 0.8125 & 0.9850 & 1.3863 \\
\bottomrule
\end{tabularx}
\end{table}

Table~\ref{tab:Control Function} reports the results of the two-stage analysis. In the first stage, we find that FinTech development is negatively associated with financial efficiency in the loan and profitability stages. It implies that FinTech companies have eroded the market share of banks in these key areas by offering lower-cost and more efficient financial services. As a result, they exert substantial competitive pressure on traditional banking operations. In the second stage, we regress $FSI$ on the FinTech-predicted MIs from the deposit, loan, and profitability stages. The results indicate that efficiency in all three stages is positively associated with FS. These findings suggest that the negative impact of FinTech on the financial efficiency of core banking operations may partly contribute to the overall adverse effect. We note, however, that this analysis is not definitive, as FinTech probably also operates through alternative channels or mechanisms that negatively affect FS of banks.

\begin{table}[H]
\centering
\small     
\caption{Mechanism analysis}
\label{tab:Control Function}
\renewcommand{\arraystretch}{0.95} 
\def\sym#1{\ifmmode^{#1}\else\(^{#1}\)\fi}
\begin{tabularx}{\textwidth}{l*{6}{>{\centering\arraybackslash}X}}
\toprule
            & (1) & (2) & (3) & (4) & (5) & (6) \\
            & $MI_d$ & $FSI$ & $MI_l$ & $FSI$ & $MI_p$ & $FSI$ \\
\midrule
$FTI$   & -0.324 &     & -0.404\sym{*} &     & -0.690\sym{*} &     \\
            & (0.305) &     & (0.209) &     & (0.358) &     \\
[1em]
$\widehat{MI_d}$ &     & 1.666\sym{***} &     &     &     &     \\
            &     & (0.628) &     &     &     &     \\
[1em]
$\widehat{MI_l}$ &     &     &     & 1.000\sym{***} &     &     \\
            &     &     &     & (0.377) &     &     \\
[1em]
$\widehat{MI_p}$ &     &     &     &     &     & 0.782\sym{***} \\
            &     &     &     &     &     & (0.295) \\
[1em]
$GDP_g$     & 0.010  & -0.011 & 0.013\sym{**} & 0.000 & -0.007 & 0.011 \\
            & (0.010) & (0.010) & (0.006) & (0.008) & (0.013) & (0.007) \\
[1em]
$FDL$        & 0.013  & -0.026 & 0.013 & -0.000 & 0.012 & -0.014 \\
            & (0.038) & (0.031) & (0.021) & (0.027) & (0.056) & (0.029) \\
[1em]
$LDR$       & 0.302  & -0.336 & -0.155 & 0.000 & 0.301 & -0.069 \\
            & (0.256) & (0.242) & (0.148) & (0.150) & (0.226) & (0.165) \\
[1em]
$NIIR$      & -0.002\sym{**} & 0.003 & -0.001 & -0.000 & -0.001 & -0.001 \\
            & (0.001) & (0.002) & (0.001) & (0.001) & (0.002) & (0.001) \\
[1em]
$ROA$       & 0.145 & -0.075 & 0.163\sym{**} & 0.000 & 0.168 & 0.034 \\
            & (0.118) & (0.117) & (0.081) & (0.097) & (0.135) & (0.089) \\
[1em]
$DAR$       & 1.045 & 1.041 & 1.477 & 0.000 & 4.688\sym{*} & -0.884 \\
            & (2.198) & (1.786) & (1.278) & (2.013) & (2.600) & (2.241) \\
[1em]
$TAS$       & -0.185 & 0.054 & -0.193\sym{**} & -0.000 & -0.559\sym{***} & 0.183 \\
            & (0.137) & (0.172) & (0.082) & (0.156) & (0.198) & (0.212) \\
[1em]
$OEX$       & 0.098 & 0.033 & 0.169\sym{***} & 0.000 & 0.411\sym{***} & -0.125 \\
            & (0.110) & (0.091) & (0.062) & (0.100) & (0.112) & (0.140) \\
[1em]
$CAR$       & 0.006 & -0.000 & 0.009 & 0.000 & 0.013 & -0.000 \\
            & (0.011) & (0.008) & (0.007) & (0.008) & (0.015) & (0.008) \\
[1em]
Constant term      & 3.288 & -3.483 & 1.898 & -0.000 & 3.905 & -1.058 \\
            & (3.186) & (2.952) & (2.295) & (2.464) & (4.851) & (2.551) \\
[1em]
Bank FE     & YES & YES & YES & YES & YES & YES \\
[1em]
Year FE     & YES & YES & YES & YES & YES & YES \\
[1em]
\midrule
Observations       & 925 & 925 & 925 & 925 & 925 & 925 \\
\(R^{2}\)   & 0.49 & 0.45 & 0.27 & 0.45 & 0.28 & 0.45 \\
\bottomrule
\multicolumn{7}{p{\textwidth}}{\footnotesize 
\textbf{Notes:} This table reports results on potential mechanisms (``channels") behind the link between FinTech and FS. The channel variables include the MIs of the deposit, loan, and profitability stages, forming the $FSI$ jointly. Variable definitions are provided in Table~\ref{tab:variable_definitions_en}. Each set of tests contains two stages of regression. In the first stage, $FTI$ is regressed on the channel variables to generate its predicted values. In the second stage, $FSI$ is regressed on the channel variable ``predicted" from the first-stage regression. $\widehat{MI_d}$, $\widehat{MI_l}$, and $\widehat{MI_p}$ denote the stage-level predicted MIs from Equation~\eqref{eq:first_stage1}. \sym{*}, \sym{**}, and \sym{***} indicate statistical significance at the 10\%, 5\%, and 1\%levels, respectively. Standard errors are clustered at the bank level and reported in parentheses. $R^2$ represents the coefficient of determination.
}
\end{tabularx}
\end{table}

\subsection{Heterogeneity analysis}
\label{subsec:Heterogeneous impact}
Based on the baseline regression, we further classify the 104 bank samples according to innovation level and marketization degree. This classification allows us to investigate how FinTech influences FS of different types of commercial banks.

Firstly, under a well-established market competition mechanism, banks face the pressure to survive the fittest. This pressure drives them to open new markets, launch new products, and invest in new technologies, thereby achieving patent-driven innovation~\citep{bos2013competition}. Against this background, the innovation level of banks is commonly measured by the number of patents they hold. Previous studies have found that banks achieved cost reduction and efficiency improvement by establishing technological barriers. Meanwhile, banks strengthen their market position through differentiated services~\citep{buchak2018fintech}. Based on these insights, banks with more patents will likely experience less significant impacts from FinTech on FS of banks.

Guided by the above analysis, we use whether the cumulative patent count of a bank exceeds the cross-sectional median as the basis for subsample grouping\footnote{The patent data is sourced from the China National Intellectual Property Administration.}. The regression results are shown in Columns (1) and (2) of Table~\ref{tab:Heterogeneity Analysis}. These results display that FinTech significantly and negatively impacts FS for banks with cumulative patent counts below the median. In contrast, the regression coefficient is insignificant for those above the median. This subsample analysis supports our earlier reasoning, i.e., the adverse effect of FinTech on FS is primarily concentrated among banks with fewer patents.

Secondly, considering the degree of marketization, we classify the banks into two groups: listed and non-listed. We do so because the listed banks typically have more diversified financing options, healthier capital positions, and more formalized risk management practices. Non-listed banks often operate under softer market constraints, relying heavily on private funding sources and facing limited external oversight. These structural advantages facilitate greater investment in technological development and help listed banks realize economies of scale~\citep{beccalli2015european}. Consequently, listed banks are likely to experience only a minor impact of FinTech development on their FS.

We estimate separate regressions for listed and non-listed banks to test this hypothesis. The results, reported in Columns (3) and (4) of Table~\ref{tab:Heterogeneity Analysis}, reveal a notable difference in how FinTech development affects the two groups. For listed banks, the estimated effect is not statistically significant, which suggests a certain degree of resilience to the disruptions brought by FinTech. Conversely, we find a significant negative relationship between FinTech development and the FS of non-listed banks. This difference reflects that banks vary in their ability to handle external shocks. Listed banks have more capital, funding sources, and transparent governance, enabling them to better handle challenges from technological changes. Non-listed banks, in comparison, often operate with thinner buffers and less institutional support, leaving them more vulnerable to the pressure of FinTech advancement.

\begin{table}[H]
\centering
\caption{Heterogeneity analysis}
\label{tab:Heterogeneity Analysis}
\def\sym#1{\ifmmode^{#1}\else\(^{#1}\)\fi}
\begin{tabularx}{\textwidth}{l*{4}{>{\centering\arraybackslash}X}}
\toprule
            & (1) & (2) & (3) & (4) \\
            & HighInnov & LowInnov & Listed & NonListed \\
            & $FSI$ & $FSI$ & $FSI$ & $FSI$ \\
\midrule
$FTI$     & -0.197         & -0.644\sym{**} & -0.509         & -0.506\sym{**} \\
            & (0.467)         & (0.261)         & (0.393)         & (0.239)         \\
[1em]
$GDP_g$       & 0.024\sym{**} & -0.000         & 0.019         & 0.000         \\
            & (0.012)         & (0.008)         & (0.012)         & (0.008)         \\
[1em]
$FDL$          & 0.083\sym{**} & -0.044         & 0.010         & -0.023         \\
            & (0.034)         & (0.035)         & (0.041)         & (0.037)         \\
[1em]
$LDR$         & -0.057         & 0.161         & 0.160         & 0.075         \\
            & (0.278)         & (0.166)         & (0.181)         & (0.175)         \\
[1em]
$NIIR$        & -0.001         & -0.002\sym{*}  & 0.001         & -0.002\sym{*}  \\
            & (0.003)         & (0.001)         & (0.003)         & (0.001)         \\
[1em]
$ROA$         & 0.142         & 0.156\sym{**} & 0.238\sym{*}  & 0.194\sym{*}  \\
            & (0.139)         & (0.074)         & (0.131)         & (0.098)         \\
[1em]
$DAR$         & 2.138         & 3.455\sym{**} & 6.627\sym{**} & 2.483         \\
            & (3.062)         & (1.597)         & (3.060)         & (1.653)         \\
[1em]
$TAS$         & -0.355\sym{**} & -0.148         & -0.654\sym{***} & -0.269\sym{**} \\
            & (0.136)         & (0.115)         & (0.236)         & (0.114)         \\
[1em]
$OEX$         & 0.464\sym{***} & 0.175\sym{***} & 0.304\sym{**} & 0.193\sym{***} \\
            & (0.094)         & (0.063)         & (0.116)         & (0.072)         \\
[1em]
$CAR$         & -0.006         & 0.019\sym{**} & 0.002         & 0.016         \\
            & (0.004)         & (0.010)         & (0.006)         & (0.010)         \\
[1em]
Constant term      & -2.036         & -0.684         & 7.103         & 2.616         \\
            & (3.001)         & (2.382)         & (5.740)         & (2.821)         \\
[1em]
Bank FE      & YES & YES & YES & YES          \\
[1em]
Year FE      & YES & YES & YES & YES           \\
[1em]
\hline
Observations       & 228         & 675         & 298         & 618         \\
\(R^{2}\)   & 0.72         & 0.45         & 0.69         & 0.38         \\
\bottomrule
\multicolumn{5}{p{\textwidth}}{\footnotesize 
\textbf{Notes:} This table presents the regression results of the heterogeneity analysis. Columns (1) and (2) show the results for subsamples with patent counts above and below the annual median patent count of all banks, respectively; Columns (3) and (4) report results for listed and non-listed banks, respectively. Differences in sample size are due to missing values in control variables. The model controls for individual and time fixed effects. Variable definitions are provided in Table~\ref{tab:variable_definitions_en}. \sym{*}, \sym{**}, and \sym{***} indicate statistical significance at the 10\%, 5\%, and 1\% levels, respectively. Standard errors are clustered at the bank level and reported in parentheses. $R^2$ represents the coefficient of determination.
}
\end{tabularx}
\end{table}

\newpage
\section{Conclusion}
\label{sec:Conclusions}
In this paper, we provide evidence that FinTech undermines the FS of commercial banks. Our main findings are as follows. First, we measure the FS of 104 Chinese commercial banks from 2015 to 2023 using a three-stage network DEA-Malmquist model. Except for the period affected by the COVID-19 pandemic, the FS of banks generally exhibits a steady upward trend. Second, empirical results across multiple model specifications, including two-way fixed effects and an IV approach, reveal that FinTech significantly diminishes the FS of commercial banks. Furthermore, the mechanism analysis reveals that the impact of FinTech mainly operates through the erosion of the loan and profitability efficiencies of banks. In addition, heterogeneity analysis indicates that banks with fewer patents and non-listed experience a greater impact from FinTech.

This paper offers an initial exploration of the impact of FinTech on FS of Chinese commercial banks, which has rarely been studied, and we contribute by bridging this gap. An open question remains as to whether the rise of FinTech will ultimately foster complementary benefits for traditional banks in the future. Some studies find that the responses of banks to FinTech can improve their performance. Incumbent banks that invest in or collaborate with FinTech firms may achieve synergies that partly offset the negative impacts of competition~\citep{hornuf2021banks}. However, other research argues that competition, interest margins, and the adverse effects of alternative digital credit might offset these gains~\citep{cuadros2023does}. This situation is particularly the case for banks lacking innovation capabilities or market discipline. Moreover, the net effect could also depend on how regulatory frameworks evolve to balance innovation with financial stability~\citep{vives2019digital}. Future research could quantify the benefits and costs of FinTech adoption for different types of banks. It also could explore how technological innovation interacts with strategic adaptation, and whether the adverse effects we document persist, attenuate, or reverse over time. Such analyses would provide valuable insights for policymakers and bank managers seeking to leverage technological development while safeguarding FS. In addition, besides FinTech, other factors might also influence the FS of banks. Future research could investigate alternative explanations from the perspectives of green finance, macroeconomic policy shocks, and internal governance structures to further enrich the relevant literature. Methodologically, future research could also consider noise-adjusted approaches, such as the NSCNLS and NStoNED models proposed by \cite{WANG2025}. These models extend network DEA to account for stochastic noise and may provide more robust efficiency estimates in multi-stage banking processes.

\section*{Acknowledgements}
This work is supported by the Young Scientists Fund of the National Natural Science Foundation of China (grant number 72301177), the Shanghai Pujiang Program (grant number 22PJC091), and the National Social Science Fund of China (grant number 24BGL062).


\bibliographystyle{apalike}
\label{sec:Reference}
\bibliography{Reference}

\end{document}